\documentstyle[epsfig,12pt,a4p]{article}

\newcommand{\pipipi}{\mbox{$\pi^+\pi^-\pi^0$ }}

\newcommand{\pisix}{\mbox{$\pi^+\pi^-\pi^+\pi^-\pi^0\pi^0$ }}
\newcommand{\pisixc}{\mbox{$\pi^+\pi^-\pi^+\pi^-\pi^+\pi^-$ }}

\newcommand{\etaeta}{\mbox{$\eta \eta$ }}
\newcommand{\etaetap}{\mbox{$\eta \eta^\prime$ }}

\newcommand{\kstkst}{\mbox{$K^*(892) \overline K^*(892)$ }}
\newcommand{\phiphi}{\mbox{$\phi \phi$ }}

\newcommand{\omom}{\mbox{$\omega \omega $ }}
\newcommand{\rhorho}{\mbox{$\rho \rho $ }}

\begin{document}
\begin{titlepage}
\def\footnoterule{\hrule width 1.0\columnwidth}
\begin{tabbing}
put this on the right hand corner using tabbing so it looks
 and neat and in \= \kill
\> {17 May 2000}
\end{tabbing}
\bigskip
\bigskip
\begin{center}{\Large  {\bf A study of the
\omom channel produced in central
pp interactions at 450 GeV/c}
}\end{center}

\bigskip
\bigskip
\begin{center}{        The WA102 Collaboration
}\end{center}\bigskip
\begin{center}{
D.\thinspace Barberis$^{  4}$,
F.G.\thinspace Binon$^{   6}$,
F.E.\thinspace Close$^{  3,4}$,
K.M.\thinspace Danielsen$^{ 11}$,
S.V.\thinspace Donskov$^{  5}$,
B.C.\thinspace Earl$^{  3}$,
D.\thinspace Evans$^{  3}$,
B.R.\thinspace French$^{  4}$,
T.\thinspace Hino$^{ 12}$,
S.\thinspace Inaba$^{   8}$,
A.\thinspace Jacholkowski$^{   4}$,
T.\thinspace Jacobsen$^{  11}$,
G.V.\thinspace Khaustov$^{  5}$,
J.B.\thinspace Kinson$^{   3}$,
A.\thinspace Kirk$^{   3}$,
A.A.\thinspace Kondashov$^{  5}$,
A.A.\thinspace Lednev$^{  5}$,
V.\thinspace Lenti$^{  4}$,
I.\thinspace Minashvili$^{   7}$,
J.P.\thinspace Peigneux$^{  1}$,
V.\thinspace Romanovsky$^{   7}$,
N.\thinspace Russakovich$^{   7}$,
A.\thinspace Semenov$^{   7}$,
P.M.\thinspace Shagin$^{  5}$,
H.\thinspace Shimizu$^{ 10}$,
A.V.\thinspace Singovsky$^{ 1,5}$,
A.\thinspace Sobol$^{   5}$,
M.\thinspace Stassinaki$^{   2}$,
J.P.\thinspace Stroot$^{  6}$,
K.\thinspace Takamatsu$^{ 9}$,
T.\thinspace Tsuru$^{   8}$,
O.\thinspace Villalobos Baillie$^{   3}$,
M.F.\thinspace Votruba$^{   3}$,
Y.\thinspace Yasu$^{   8}$.
}\end{center}

\begin{center}{\bf {{\bf Abstract}}}\end{center}

{
The reaction
$ pp \rightarrow p_{f} (\omega \omega ) p_{s}$ has been studied at 450 GeV/c
and a spin analysis of the \omom channel has been performed
for the first time in central production.
Evidence is found for the $f_2(1910)$ in the $J^{PC}$~=~$2^{++}$
wave with spin projection $J_Z$~=~2.
This is the only state observed in central production
with spin projection $J_Z$~=~2.
Its $dP_T$ and $\phi$ dependencies are similar to those
observed for other glueball candidates.
In addition, evidence is found for a state with $J^{PC}$~=~$4^{++}$
consistent with the $f_4(2300)$.
The $f_0(2000)$, previously observed in the
$\rho \rho$ final state, is confirmed.
}
\bigskip
\bigskip
\bigskip
\bigskip\begin{center}{{Submitted to Physics Letters}}
\end{center}
\bigskip
\bigskip
\begin{tabbing}
aba \=   \kill
$^1$ \> \small
LAPP-IN2P3, Annecy, France. \\
$^2$ \> \small
Athens University, Physics Department, Athens, Greece. \\
$^3$ \> \small
School of Physics and Astronomy, University of Birmingham, Birmingham, U.K. \\
$^4$ \> \small
CERN - European Organization for Nuclear Research, Geneva, Switzerland. \\
$^5$ \> \small
IHEP, Protvino, Russia. \\
$^6$ \> \small
IISN, Belgium. \\
$^7$ \> \small
JINR, Dubna, Russia. \\
$^8$ \> \small
High Energy Accelerator Research Organization (KEK), Tsukuba, Ibaraki 305-0801,
Japan. \\
$^{9}$ \> \small
Faculty of Engineering, Miyazaki University, Miyazaki 889-2192, Japan. \\
$^{10}$ \> \small
RCNP, Osaka University, Ibaraki, Osaka 567-0047, Japan. \\
$^{11}$ \> \small
Oslo University, Oslo, Norway. \\
$^{12}$ \> \small
Faculty of Science, Tohoku University, Aoba-ku, Sendai 980-8577, Japan. \\
\end{tabbing}
\end{titlepage}
\setcounter{page}{2}
\bigskip
\par
The \omom channel has been studied in several different production mechanisms.
In $\pi^-p$ interactions the \omom final state has been studied by the
NA12~\cite{na12omom} and VES~\cite{vesomom} collaborations. In both
experiments clear signals
were observed at 1.6 and 1.9 GeV and were found to
have $J^{PC}$~=~$2^{++}$, called the $f_2(1640)$ and $X(1910)$~\cite{PDG98}.
In addition, the VES collaboration reported evidence for an \omom decay
mode of the $f_4(2050)$ and, more recently, for a $J^{PC}$~=~$4^{++}$
object in the 2.3 GeV region~\cite{vesfrascati}.
In $p \bar{p}$ annihilations
C. Baker et al., using the
data from the Crystal Barrel experiment~\cite{cbomom},
have reported evidence for a structure similar to the $f_2(1640)$
in the \omom final state
but have shown that this state can be interpreted as being due to the
$f_2(1565)$ previously observed in the $\pi \pi$ final
state~\cite{PDG98}.
The PDG~\cite{PDG98} lists the $X(1910)$ observed in the \omom
final state with another $J^{PC}$~=~$2^{++}$ resonance with similar
mass and width observed in the \etaetap final state.
In central production the WA102 experiment did not
observe the $f_2(1565)$ in the
$\pi \pi$ final state~\cite{pipipap}, therefore,
the centrally produced \omom channel can give information
on the validity of the $f_2(1565)/f_2(1640)$ assignment.
In addition, in the \etaetap
final state of the WA102 experiment~\cite{etaetappap}
a peak was observed at 1.9~GeV which was consistent in mass and width
with the $X(1910)$.
A spin analysis showed that this state was consistent with having
$J^{PC}$~=~$1^{-+}$ with spin projection $J_Z$~=~1 or
$J^{PC}$~=~$2^{++}$ with spin projection $J_Z$~=~2.
If the latter hypothesis were true then
this was the first time that a state had been
observed in central production that was
produced with spin projection
$J_Z$~=~2.
Hence, if the states observed in \omom and \etaetap are the same
and the $X(1910)$ has $J^{PC}$~=~$2^{++}$,
the $X(1910)$ should be observed in the $J_Z$~=~2 projection
in the \omom final state.
In central production,
the \omom final state was previously observed by the
WA76 experiment~\cite{wa76omom}
but only 80 events were observed and hence no strong conclusions
could be drawn.
\par
In this paper, a study is presented of the \omom
final state formed in the reaction
\begin{equation}
pp \rightarrow p_{f} (\omega \omega) p_{s}
\label{eq:e}
\end{equation}
at 450~GeV/c.
It represents more than a factor of 60 increase in statistics over previous
data on the centrally produced \omom final state~\cite{wa76omom}
and, moreover, will present a spin analysis of this
channel in central production.
The data come from the WA102 experiment
which has been performed using the CERN Omega Spectrometer,
the layout of which is
described in ref.~\cite{WADPT}.
Reaction~(\ref{eq:e})
has been isolated using the \pipipi decay mode of both $\omega$s.
The reaction
\[
pp \rightarrow p_{f} (\pi^+ \pi^-\pi^+\pi^-\pi^0 \pi^0) p_{s}
\]
has been isolated
from the sample of events having six
outgoing
charged tracks and four $\gamma$s reconstructed in the GAMS-4000
calorimeter,
by first imposing the following cuts on the components of the
missing momentum:
$|$missing~$P_{x}| <  17.0$ GeV/c,
$|$missing~$P_{y}| <  0.16$ GeV/c and
$|$missing~$P_{z}| <  0.12$ GeV/c,
where the $x$ axis is along the beam
direction.
The two photon mass spectrum,
when
the mass of the other $2\gamma$-pair lies
within a band around the $\pi^0$ mass (100--170 MeV), shows
a clear $\pi^0$ signal with small background.
Events containing a fast $\Delta^{++}(1232) $
were removed if $M(p_{f} \pi^{+}) < 1.3 $ GeV, which left
294 463 centrally produced \pisix events.
\par
Fig.~\ref{fi:1}a) shows a lego plot of $M(\pi^+\pi^-\pi^0)$ versus
$M(\pi^+\pi^-\pi^0)$ (four combinations per event).
A clear signal of the \omom channel can be observed.
Fig.~\ref{fi:1}b) shows the \pipipi mass spectrum if
the other \pipipi combination is compatible with being an $\omega$
(0.76~$\leq$~M(\pipipi)~$\leq$~0.81~GeV) where a clear $\omega$ signal
can be observed.
A tight cut has been used around the $\omega$ signal to increase the signal
to background ratio in the selected sample.
In order to decrease the background further
the parameter $\lambda$ is introduced which describes
the $\omega$ decay on the Dalitz plot and
is defined as:
\[
\lambda = \frac{|\vec{p}_+ \times \vec{p}_-|^2}
{\frac{3}{4}(\frac{1}{9}m^2-m_\pi^2)^2}
\]
where $|\vec{p}_+ \times \vec{p}_-|$
is proportional to the decay matrix element for
$\omega \rightarrow \pi^+\pi^-\pi^0$, $\vec{p}_\pm$ is the
three momentum of the $\pi^\pm$ in the $\omega$ rest frame and
$m^2$ is the \pipipi effective mass squared.
Superimposed on
fig.~\ref{fi:1}b) as a shaded histogram
is the \pipipi mass distribution for $\lambda$~$>$~0.3. As can be seen
the signal to background ratio in the $\omega$ region has increased.
\par
The \omom final state has been selected using the \pipipi
mass cuts described above and by requiring that the
$\lambda$~$>$~0.3 for each $\omega$ candidate.
The resulting \omom mass spectrum is shown in fig.~\ref{fi:1}c) and consists
of 5067 events.
As can be seen there is a peak in the 1.9~GeV region.
\par
The background below the $\omega$ signal has several sources
including combinatorics and other channels.
The combinatorial background is removed, in part,
in the selection procedure.
The
remaining background is approximately 27 \%.
Four methods have been used to determine the effects of this background;
studying the side bands around the $\omega$ signal, studying events that
do not balance momentum, studying events that do not
pass the $\lambda$ selection cuts and studying events from the
\pisixc channel.
Since the majority of the background is due to other
physical channels for example $a_1(1260) a_1(1260)$ or $\omega a_1(1260)$
production, the two methods that best reproduce the background are the
one using events that do not pass the $\lambda$ cut
and the other uses events from the \pisixc channel. These two methods give
a very similar representation of the background.
In the remainder of this paper the method used to determine
the background will be the mean of these two methods.
Superimposed on the \omom mass spectrum in fig~\ref{fi:1}c)
as a shaded histogram is the
estimate of the background.
\par
A spin analysis of the centrally produced \omom system
has been performed using the method
described in ref.~\cite{re:wa914pi} for the $\rho\rho$ final state
modified for the \omom channel.
The z axis is defined by the momentum vector of the
exchanged particle with the greatest four-momentum transferred
in the \omom centre of mass.
Assuming that
only angular momenta up to 4 contribute,
the amplitudes have been calculated in the spin-orbit (LS)
scheme using spherical
harmonics.
\par
In order to perform a spin parity analysis the
log likelihood function, ${\cal L}_{j}=\sum_{i}\log P_{j}(i)$,
is defined by combining the probabilities of all events in 50 MeV
\omom mass bins from 1.5 to 3.0 GeV.
The incoherent sum of various
event fractions $a_{j}$ is calculated
so as to include more than one wave in the fit,
\begin{equation}
{\cal L}=\sum_{i}\log \left(\sum_{j}a_{j}P_{j}(i) +
(1-\sum_{j}a_{j})\right)
\end{equation}
where the term
$(1-\sum_{j}a_{j})$ represents the phase space background.
The negative log likelihood function ($-{\cal L} $) is then minimised using
MINUIT~\cite{re:MINUIT}. Coherence between different $J^{P}$ states
has
been neglected in the fit.
Different combinations of waves have been tried and
insignificant contributions have been removed from the final fit.
\par
It is found necessary to introduce the $J^{PC}$~=~$2^{++}$ wave
with both $J_Z$~=~0 and 2, the $J^{PC}$~=~$0^{++}$ wave and
the $J^{PC}$~=~$4^{++}$ wave with $J_Z$~=~1.
The results of the best fit are shown in
fig.~\ref{fi:2}.
\par
The $J^{PC}$~=~$2^{++}$ wave with $J_Z$~=~2 shows a peak at 1.9 GeV.
This wave has been fitted using a spin 2 relativistic Breit-Wigner
and a linear background and
is shown superimposed.
The fit gives M~=~1897~$\pm$~11~MeV, $\Gamma$~=~202~$\pm$~32~MeV,
parameters consistent with those of the $X(1910)$
found from a fit to the \etaetap final state~\cite{etaetappap}.
Hence this consistent with the fact that the $X(1910)$ has \omom and \etaetap
decay modes,
and we shall refer to it as the $f_2(1910)$ hereafter.
Correcting for the unseen decay modes and the effects of the detector,
the branching ratio \omom/\etaetap
of the $f_2(1910)$ is
2.6~$\pm$~0.6.
There was no evidence for any wave with $J_Z$~=~2 in the \etaeta
final state of the WA102 experiment~\cite{etaetapap} and hence an upper
limit for the
branching ratio \etaeta/\etaetap
of the $f_2(1910)$ has been calculated to
be $<$~0.2~($90\; \%$ CL).
\par
The $J^{PC}$~=~$2^{++}$ wave with $J_Z$~=~0 shows a broad enhancement.
Superimposed on the wave is a shaded histogram
representing the $f_2(1640)$.
As can be seen the
$J^{PC}$~=~$2^{++}$ wave with $J_Z$~=~0
is not compatible with the $f_2(1640)$ observed
by other experiments. This non observation does not contradict the
claim that the $f_2(1640)$ is an \omom decay mode of the
$f_2(1565)$ since this state is also not observed in
central production.
\par
The $J^{PC}$~=~$0^{++}$ wave  shows some activity near threshold
and a broad enhancement around 2~GeV.
In a previous analysis of the 4$\pi$ channel, the
WA102 experiment observed
a similar structure
in the $J^{PC}$~=~$0^{++}$ $\rho \rho$ wave
which was identified with
the $f_0(2000)$.
Superimposed on the wave is a shaded histogram
representing the $f_0(2000)$
assuming that the branching ratio $\rho\rho$/\omom~=~3
as expected for a isoscalar resonance.
This well
represents the wave in the 2~GeV region.
\par
The $J^{PC}$~=~$4^{++}$ wave with $J_Z$~=~1 shows
no evidence for the $f_4(2050)$ but does show
a peak at 2.3~GeV.
The change in
log likelihood
in the three 50 MeV bins around the 2.3~GeV peak
produced by introducing the
$J^{PC}$~=~$4^{++}$ wave with $J_Z$~=~1 is
$\Delta{\cal L} = 28$.
This is the first time that  it has been found necessary to
introduce any wave with $J$~$>$~2 in the WA102 data.
This wave has been fitted using a spin 4 relativistic Breit-Wigner
and a linear background and is superimposed on the wave.
The fit gives M~=~2332~$\pm$~15~MeV, $\Gamma$~=~260~$\pm$~57~MeV
parameters consistent with those
found by the VES experiment~\cite{vesfrascati}.
This state is most likely the $f_4(2300)$ observed previously
in other experiments~\cite{f42300} and we shall refer to it as so hereafter.
\par
States that have a decay to \omom might also be expected to
have a decay to \rhorho. As was observed above there appears
to be evidence for an \omom decay of the $f_0(2000)$ previously
observed in the \rhorho final state.
In the previous analysis of the 4$\pi$ final state~\cite{pi4pap},
no evidence was claimed for either a
$J^{PC}$~=~$2^{++}$ \rhorho wave with $J_Z$~=~2 or a
$J^{PC}$~=~$4^{++}$ \rhorho wave with $J_Z$~=~1.
Because of the large number of possible waves in the 4$\pi$
final state ( $\sim$ 180 for $J$~$\le$~2)
only waves that
changed the
log likelihood by more than 100 were considered.
The
$J^{PC}$~=~$2^{++}$ \rhorho wave with $J_Z$~=~2 was rejected
because it changed the likelihood by $\sim$ 60.
The $J^{PC}$~=~$4^{++}$ \rhorho wave with $J_Z$~=~1 was not
considered because only waves with $J$~$\le$~2 were included in the fit.
If the
$J^{PC}$~=~$2^{++}$ \rhorho wave with $J_Z$~=~2 and the
$J^{PC}$~=~$4^{++}$ \rhorho wave with $J_Z$~=~1
are both introduced into the fit of the $\pi^+\pi^-\pi^+\pi^-$ channel
then the log likelihood increases by 58 units in the region of the
$f_2(1910)$ and 27 units in the $f_4(2300)$ region.
If the signal in the
$J^{PC}$~=~$2^{++}$ \rhorho wave with $J_Z$~=~2 is interpreted as being due
to the $f_2(1910)$ then
after correcting for the unseen decay modes and the effects of the detector
the branching ratio \rhorho/\omom
of the $f_2(1910)$ is
2.6~$\pm$~0.4 consistent with it being at isoscalar resonance.
Similarly
if the signal in the
$J^{PC}$~=~$4^{++}$ \rhorho wave with $J_Z$~=~1 is interpreted as being due
the $f_4(2300)$ then after
correcting for the unseen decay modes and the effects of the detector
the branching ratio \rhorho/\omom
of the $f_4(2300)$ is
2.8~$\pm$~0.5.
\par
In previous analyses a study has been made of how different
resonances are produced as a function of the
parameter $dP_T$, which is the difference
in the transverse momentum vectors of the two exchange
particles~\cite{WADPT,closeak}, and as a function of
the azimuthal angle $\phi$ which is defined as the angle between the $p_T$
vectors of the two outgoing protons.
A study of the background subtracted \omom system over the whole
mass range as a function of
$dP_T$
has been performed.
The fraction of all \omom
production has been calculated for
$dP_T$$\leq$0.2 GeV, 0.2$\leq$$dP_T$$\leq$0.5 GeV and $dP_T$$\geq$0.5 GeV and
gives
0.12~$\pm$~0.02, 0.36~$\pm$~0.02 and 0.52~$\pm$~0.02 respectively.
This results in a ratio of production at small $dP_T$ to large $dP_T$ of
0.23~$\pm$~0.04.
This ratio is much lower than has been observed~\cite{kstkstpap,phiphipap}
in the \kstkst and \phiphi final states. However, the latter final states
have been shown to be dominantly due to the $f_2(1950)$
which is produced mainly at small $dP_T$~\cite{pi4papr}.
\par
The amount of $f_2(1910)$ has also been determined in the same
$dP_T$ intervals and gives
0.20~$\pm$~0.04, 0.62~$\pm$~0.07 and 0.18~$\pm$~0.04 respectively.
This results in a ratio of production at small $dP_T$ to large $dP_T$ of
1.1~$\pm$~0.3.
This value is consistent with what has been observed for
the glueball candidates the
$f_0(1500)$, $f_0(1710)$ and $f_2(1950)$~\cite{pi4papr,pipikkpap,memoriam}.
\par
The azimuthal angle ($\phi$) between the $p_T$
vectors of the two protons
is shown in
fig.~\ref{fi:4}a) for the background subtracted \omom channel for the
entire mass range and in
fig.~\ref{fi:4}b) for the $f_2(1910)$.
The distribution for the $f_2(1910)$ is similar to that observed for
other glueball candidates~\cite{pi4papr,pipikkpap}.
\par
In summary,
a spin analysis of the \omom channel has been performed
for the first time in central production.
Evidence is found for the $f_2(1910)$ in the $J^{PC}$~=~$2^{++}$
wave with spin projection $J_Z$~=~2.
This is the only state observed in central production
with spin projection $J_Z$~=~2.
Its $dP_T$ and $\phi$ dependencies are similar to those
observed for other glueball candidates.
In addition, evidence is found for a state with $J^{PC}$~=~$4^{++}$
consistent with the $f_4(2300)$.
The $f_0(2000)$, previously observed in the
$\rho \rho$ final state, is confirmed.
\begin{center}
{\bf Acknowledgements}
\end{center}
\par
This work is supported, in part, by grants from
the British Particle Physics and Astronomy Research Council,
the British Royal Society,
the Ministry of Education, Science, Sports and Culture of Japan
(grants no. 07044098 and 1004100), the French Programme International
de Cooperation Scientifique (grant no. 576)
and
the Russian Foundation for Basic Research
(grants 96-15-96633 and 98-02-22032).

\clearpage
{ \large \bf Figures \rm}
\begin{figure}[h]
\caption{
Selection of the \omom final state:
a) M(\pipipi) versus M(\pipipi),
b) M(\pipipi) if the other \pipipi combination
is in the $\omega$ band
(0.76~$\leq$~M(\pipipi)~$\leq$~0.81~GeV).
Superimposed as a shaded histogram
is the case for $\lambda$~$>$~0.3.
c) The \omom mass spectrum.
Superimposed as a shaded histogram
is an estimation of the background contribution.
}
\label{fi:1}
\end{figure}
\begin{figure}[h]
\caption{
Results of the spin analysis for the \omom channel:
a) The total mass spectrum,
b)~$2^{++}$~$J_Z=2$,
c)~$2^{++}$~$J_Z=0$,
d)~$0^{++}$ and
e)~$4^{++}$~$J_Z=1$.
The superimposed curves are the resonance contributions coming from
the fits described in the text.}
\label{fi:2}
\end{figure}
\begin{figure}[h]
\caption{
The $\phi$ distribution for the a) the \omom channel and b) for the
$f_2(1910)$.
}
\label{fi:4}
\end{figure}
\newpage
\begin{center}
\epsfig{figure=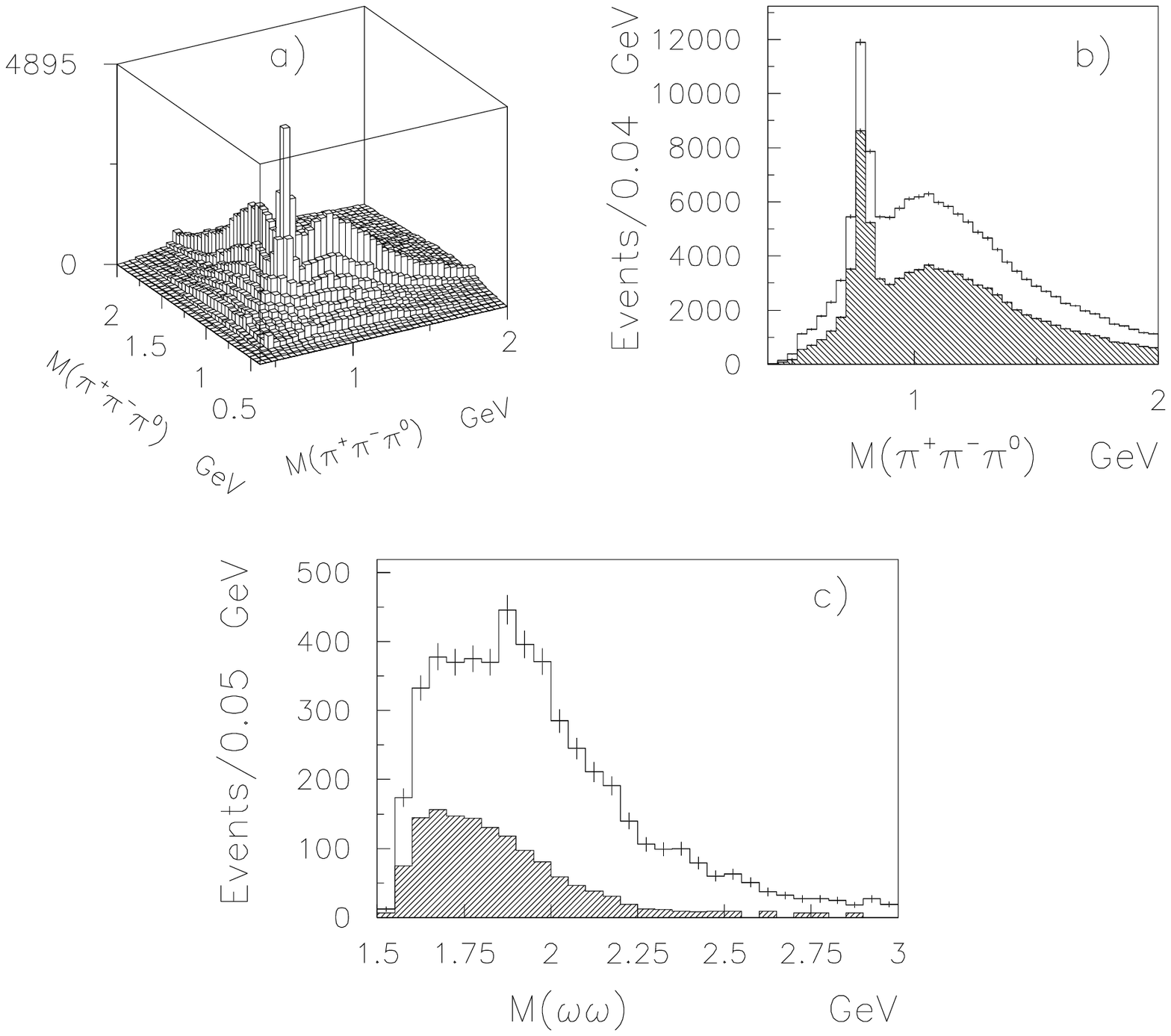,height=22cm,width=17cm}
\end{center}
\begin{center} {Figure 1} \end{center}
\newpage
\begin{center}
\epsfig{figure=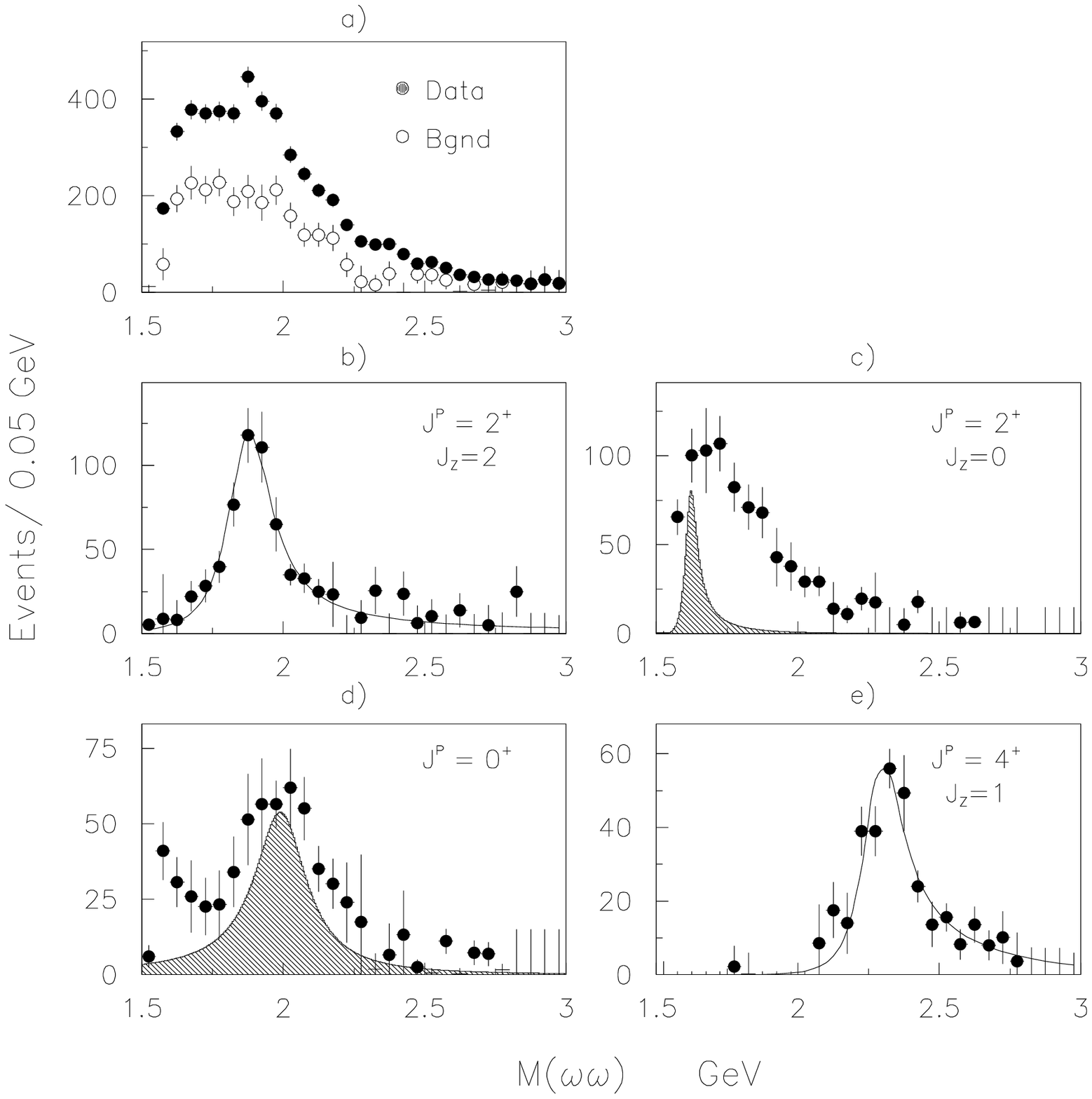,height=22cm,width=17cm}
\end{center}
\begin{center} {Figure 2} \end{center}
\newpage
\begin{center}
\epsfig{figure=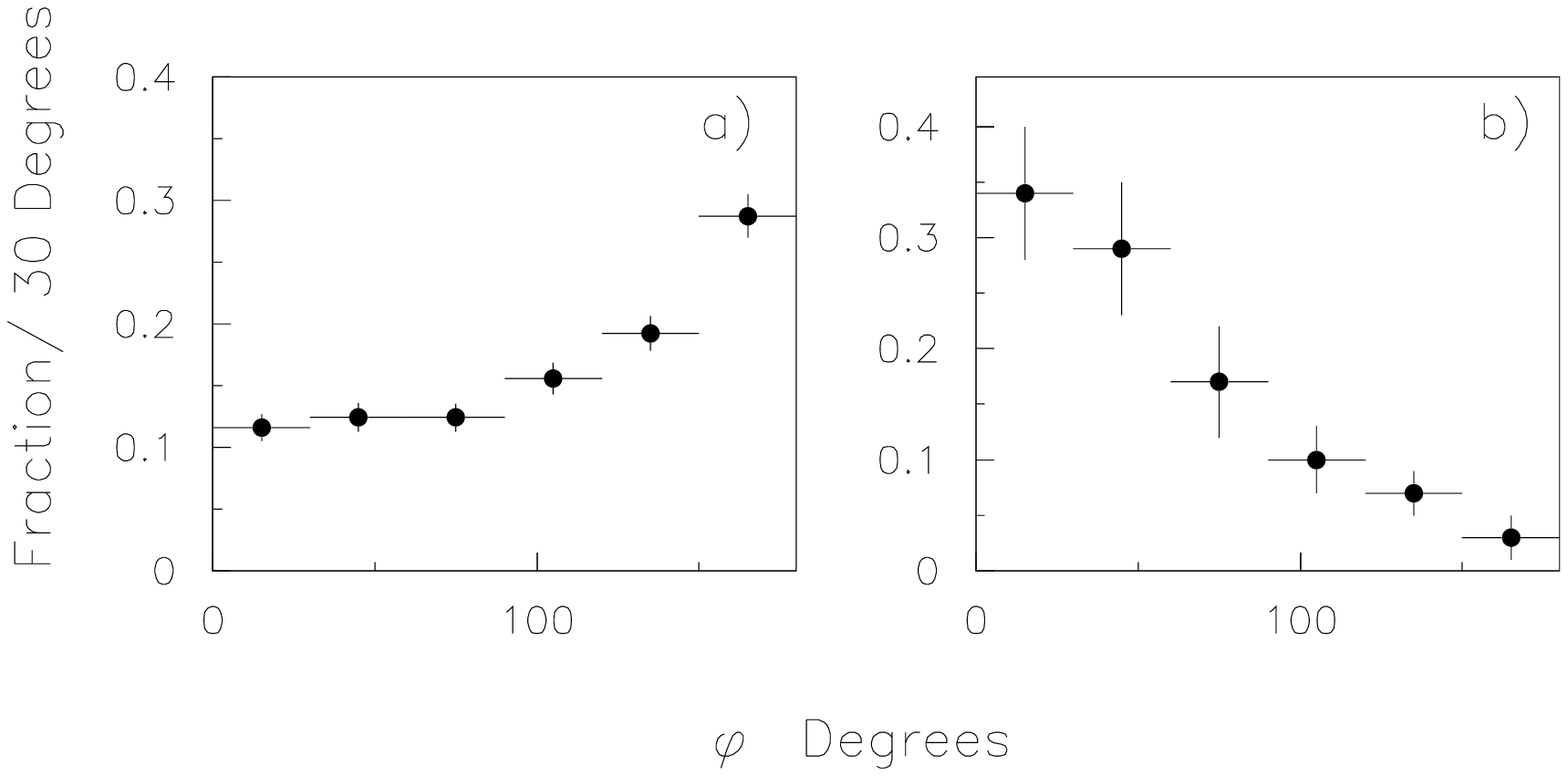,height=22cm,width=17cm}
\end{center}
\begin{center} {Figure 3} \end{center}
\end{document}